\documentclass[useAMS,usenatbib]{mn2e}
\usepackage{epsfig}
\usepackage{graphicx}
\usepackage{amsmath}
\usepackage{subfig}
\usepackage{cleveref} 
\usepackage{color}

\newcommand{\ba}{\begin{eqnarray}}
\newcommand{\ea}{\end{eqnarray}}
\newcommand{\be}{\begin{equation}}
\newcommand{\ee}{\end{equation}}
\newcommand{\bfv}{{\bf v}}
\newcommand{\bu}{{\bf u}}

\newcommand{\tomega}{{\tilde\omega}}
\newcommand{\rmRe}{{\rm Re}}

\def\go{\mathrel{\raise.3ex\hbox{$>$}\mkern-14mu
             \lower0.6ex\hbox{$\sim$}}}
\def\lo{\mathrel{\raise.3ex\hbox{$<$}\mkern-14mu
             \lower0.6ex\hbox{$\sim$}}}
\newcommand{\aap}{A\&A}
\newcommand{\apj}{ApJ}
\newcommand{\apjl}{ApJL}
\newcommand{\mnras}{MNRAS}
\newcommand{\pasj}{PASJ}
\newcommand{\icarus}{Icarus}

\begin{document}

\title[3D Rossby Wave Instability]
{Rossby Wave Instability in three dimensional discs}
\author[H. Meheut, C. Yu and D. Lai]
{Heloise Meheut$^1$\thanks{Email:meheut@space.unibe.ch},
Cong Yu$^{2,3,4}$ and Dong Lai$^2$\\
$^1$ Physikalisches Institut \& Center for Space and Habitability, Universit\"at Bern, 3012 Bern, Switzerland\\
$^2$ Department of Astronomy, Cornell University, Ithaca, NY 14853, USA\\
$^3$ National Astronomical Observatories/Yunnan Astronomical Observatory, Chinese Academy of Sciences, Kunming, 650011, China\\
$^4$ Key Laboratory for the Structure and Evolution of Celestial Objects, Chinese Academy of Sciences, Kunming, 650011, China}
\pagerange{\pageref{firstpage}--\pageref{lastpage}} \pubyear{2011}

\label{firstpage}
\maketitle
\begin{abstract}
The Rossby wave instability (RWI) is a promising mechanism for producing large-scale vortices in protoplanetary discs. The instability operates around a density bump in the disc, and the resulting vortices may facilitate planetesimal formation and angular momentum transfer in the disc dead zone. Most previous works on the RWI deal with two-dimensional (height-integrated) discs. However, vortices may have different dynamical behaviours in 3D than in 2D.  Recent numerical simulations of the RWI in 3D global discs by Meheut et al. have revealed intriguing vertical structure of the vortices, including appreciable vertical velocities.  In this paper we present a linear analysis of the RWI in 3D global models of isothermal discs. We calculate the growth rates of the Rossby modes (of various azimuthal wave numbers $m=2-6$) trapped around the fiducial density bump and carry out 3D numerical simulations to compare with our linear results.  We show that the 3D RWI growth rates are only slightly smaller than the 2D growth rates, and the velocity structures seen in the numerical simulations during the linear phase are in agreement with the velocity eigenfunctions obtained in our linear calculations. This numerical benchmark shows that numerical simulations can accurately describe the instability. The angular momentum transfer rate associated with Rossby vortices is also studied.
\end{abstract}


\section{Introduction}
 
Vortices may play an importance role in protoplanetary discs and
planet formation. First, long-lived anticyclonic vortices concentrate
dust grains in their centers and accelerate the growth of meter-sized
solids \citep{BAR95,TBD96,BRA99,GL00,JAB04,HK10}, leading to the formation of planetesimals. Second, vortices may generate disordered (turbulent) flow, which induces angular momentum transfer in the radial direction. This is particularly relevant to accretion in the disc dead zone (e.g. \citealt{G96,T08}), where the gas is not ionised by stellar radiation nor by cosmic rays, and turbulence driven by magnetorotational instability is ineffective. Finally, vortices can also form at the edge of planetary gaps \citep{dAD07,YLL10,LP11,LP211}, thereby affecting the rate of planet migration. In these different contexts, but mainly for the concentration of solids particles, it is important to understand the 3D velocity structure of the vortices.

A promising mechanism for producing vortices is the Rossby Wave Instability
(RWI) (\citealt{LOV99} \citealt{LI00,LI01}).
The RWI is fundamentally related to Kelvin-Helmholtz instability 
of shearing flows and Rayleigh's inflection point theorem
(e.g., \citealt{PAP85,PAP89,LIT09}). 
For two-dimensional (vertically integrated) barotropic discs, 
the RWI relies on the existence of an extremum in the background
fluid vortensity, defined by 
\be
\zeta={\frac{(\nabla\times {\bf v})\cdot {\hat z}}{\Sigma}}
={\frac{\kappa^2}{2\Omega\Sigma}}, 
\ee 
where $\Sigma$ is the disc surface density, $\bf v$ the fluid velocity,
$\Omega$ the disc rotation rate and 
\be 
\kappa^2={\frac{2\Omega}{r}}{\frac{d}{dr}}(r^2\Omega)
\label{eq_kappa2}
\ee 
is the square of the radial epicyclic frequency (so that $\kappa^2/2\Omega$ is the vorticity).
Since Rossby waves propagate along the gradient of vortensity, the instability can be understood as arising from the interaction between two Rossby waves standing on each side of the vortensity extremum.
In protoplanetray discs, the vortensity extremum is expected to form at the boundaries of the dead zone, as the difference in accretion rate in these regions forms a bump in the density profile \citep{VBF06,INB06,T08,KLG09}. Recent MHD simulations \citep{DFT10} are beginning to address various complex physics related to the formation of this density/pressure bump.

So far most of the studies of the RWI have considered two-dimensional (height-integrated) discs (e.g. \citealt{LJK09,TAM06,VAR06,RJS11}). However, while 2D vortices are long-lived (but see \citealt{CO10}), the situation is not clear in three-dimensions (see \citealt{BAM05,SHE06,LES09}), and it is important to study the formation of vortices and their stability in 3D discs. On the analytical side, \citet{U10} used local shallow-water approximation to examine the RWI in idealized situations where the density bump is replaced by step functions. \citet{LIT09} considered vertically unstratified discs in shearing box approximation.
Some 3D aspects of other instabilities controlled by corotation, such as the Papaloizou-Pringle instability in rotating tori, have also been studied in previous works (e.g. \citealt{PAP85,GGN86,LAB09}).
However, the 3D aspect of the RWI, where Rossby waves are self trapped around the density bump, has not been studied.

Recently, Meheut et al. (2010, 2012) \nocite{MEH10,MKC12} carried out global 3D numerical simulations of the RWI in a full 3D disc. The simulations showed that the RWI can develop in 3D discs as in 2D, but the resulting Rossby vortices have significant vertical structure, contrary to what was expected.
In particular, the vertical velocity plays an important role in the long-term evolution of the vortices (Meheut et al., 2012). Indeed, the elliptical instability \citep{KER02,LES09} that is responsible for the decay of unstratified (or with no vertical flow) vortices is a 3D mechanism, and may be affected by the vertical structure. Moreover, the vertical velocity can modify the concentration of solids inside the vortices, thereby influencing the formation of planetesimals.

The goal of this paper is to provide an understanding of the 3D structure of Rossby vortices that has been observed in the 3D simulations of the RWI. To this end, we carry out global linear analysis of the RWI in full 3D discs and compare our results with those obtained from numerical simulations. Our paper is organized as follows. We first present the 3D linear perturbation equations (Section \ref{equations}) and 
the numerical methods for solving the eigenvalue problem (Section \ref{methods}). The results of our linear calculations are described and discussed in Section \ref{results}. In order to compare with the results of Meheut et al. (2010), we present in Section \ref{simulations} full numerical simulations of the development of Rossby vortices using the same setup and parameters as our linear analysis.
We conclude in Section 6.

\section{Equations}\label{equations}

The linear perturbation equations for stratified 3D discs are similar to those given in \citet{ZL06} in their study of the tidal excitation of 3D waves in discs (see also \citealt{OKF87,TTW02}). Here we summarize the notations and key equations relevant to our analysis.

\subsection{Governing equations}

We consider a geometrically thin gas disc and adopt cylindrical coordinates $(r,\varphi,z)$. The governing equations read:
\ba
&&\rho{\partial \bfv\over\partial t}+(\bfv\cdot\nabla)\rho\bfv
=-\nabla p -\rho \nabla \Phi_G\\
&& {\partial\rho\over\partial
t}+\nabla\cdot(\rho\bfv)=0,
\ea
where $\rho$ is the density, $p$ the pressure, $\bfv$ the velocity.
The disc is assumed to be non-self-gravitating, the gravitational potential $\Phi_G$ is the due to the central star only. We will consider an isothermal equation of state, $p=c_s^2\rho$, with $c_s$ the constant sound speed.

The unperturbed disc has velocity $\bfv_e=(0,r\Omega,0)$, where the angular velocity $\Omega=\Omega(r)$ is taken to be a function of $r$ alone.

Thus the vertical density profile is given by 
\be 
\rho_e(r,z)={\Sigma\over\sqrt{2\pi}h}\exp (-Z^2/2),\quad
{\rm with}~~ Z=z/h 
\label{eq:rho0}\ee 
where $h=h(r)=c_s/\Omega_\perp$ is the disc scale height, 
$\Sigma=\Sigma(r)=\int dz\,\rho_e$ is the surface density, and $\Omega_\perp$ is the vertical oscillation frequency of the disc and is equal to the Keplerian frequency $\Omega_K=(GM/r^3)^{1/2}$ where $G$ is the gravitational constant and $M$ the central star mass.

\subsection{Linear pertubations equations}

We now consider linear perturbation of the disc. For simplicity, we shall assume that the perturbation is isothermal, so that the (Eulerian) density and pressure perturbations are related by $\delta p=c_s^2\delta\rho$. The linear perturbation equations read
\ba
&&{\partial \bu\over\partial t}+(\bfv_e\cdot\nabla)\bu+(\bu\cdot\nabla)\bfv_e
=-\nabla\eta,\label{eq:u}\\
&& {\partial\delta\rho\over\partial
t}+\nabla\cdot(\rho_e\bu+\bfv_e\delta\rho)=0, 
\label{eq:rho}\ea 
where $\bu=\delta\bfv$ is the (Eulerian) velocity perturbation, and $\eta\equiv\delta p/\rho_e$ is the enthalpy perturbation. Without loss of generality, the perturbation variables are assumed to depend on $t$ and $\varphi$ as
\be
\bu,\eta,\delta\rho \propto \exp(im\varphi-i\omega t),
\ee
where the azimuthal mode number $m>0$ is an integer and $\omega$ is the (complex) wave frequency.
Equations (\ref{eq:u}) and (\ref{eq:rho}) then reduce to
\ba
&&\!\!\!-i\tomega u_r-2\Omega u_\varphi=-{\partial\over\partial r}\eta,
\label{eq:fluid1}\\
&&\!\!\!
-i\tomega u_\varphi +{\kappa^2\over 2\Omega}u_r=-{im\over r}\eta,
\label{eq:fluid22}\\
&&\!\!\!-i\tomega u_z=-{\partial \over\partial z}\eta,\label{eq:fluid3}\\
&&\!\!\!-i\tomega {\rho_e\over c_s^2}\eta+{1\over r}{\partial\over\partial
r} (r\rho_e u_r)+{im\over r}\rho_e u_\varphi +{\partial\over\partial
z}(\rho_e u_z)=0. \label{eq:fluid}\ea 
Here $\tomega$ is the ``Doppler-shifted'' frequency
\be
\tomega=\omega-m\Omega,
\ee
and $\kappa$ is the radial epicyclic frequency defined by equation~(2).

We consider radiative boundary condition such that waves propagate away from the density bump at both the inner and outer boundaries of the disk (e.g., \citealt{YL09}).
They are defined as follows. We take
\begin{equation}
\frac{d u_{r0,2}}{d r} = i k u_{r0,2 }\ , \ \frac{d \delta
h_{0,2}}{d r} = i k \delta h_{0,2} \ ,
\end{equation}
we then substitute these equations into the disk perturbation equations, we will arrive at a complex polynomial at the boundary that determines the value of $k$. Then we choose the appropriate root of the polynomial so that the group velocity is in the outgoing direction. We also implement the null radial velocity boundary condition to check how the eigenvalues are affected by different boundary conditions.\\

To separate out the $z$-dependence, we expand the perturbations with Hermite polynomials $H_n$
(see \citealt{OKF87,K01,O08})
\ba
\left[\begin{array}{c}
\eta(r,z)\\
u_r(r,z)\\
u_\varphi(r,z)\end{array}\right]
&=& \sum_n \left[\begin{array}{c}
\eta_n(r)\\
u_{rn}(r)\\
u_{\varphi n}(r)\end{array}\right] H_n(Z),\nonumber\\
u_z(r,z) &=& \sum_n u_{zn}(r)\,H_n'(Z),
\label{eq:expand}\ea 
where $H'_n=dH_n/dZ$, and the Hermite polynomial is defined by $H_n(Z)\equiv(-1)^n e^{Z^2/2}d^n(e^{-Z^2/2})/dZ^n$. Note that $H_0=1$, $H_1=Z$ and $H_2=Z^2-1$.

\begin{figure}
\begin{center}
\includegraphics[width=0.4\textwidth]{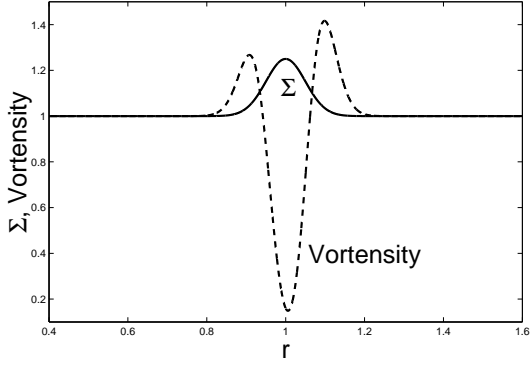}
\end{center}
\caption{Radial profile of the surface density and vortensity $\zeta$. 
The radius is in units of $r_0$, and the surface density is normalised 
to $\Sigma_0$.}
\label{fig:dens}
\end{figure}

With the expansion in (\ref{eq:expand}), the fluid equations (\ref{eq:fluid1})-(\ref{eq:fluid})
become
\ba 
&&-i\tomega u_{rn}-2\Omega u_{\varphi n}=
-{d\over dr} \eta_n+{n\mu\over r}\eta_n\nonumber\\
&& \qquad \qquad \qquad +{(n+1)(n+2)\mu\over r}\eta_{n+2},\label{u1}\\
&&-i\tomega u_{\varphi n} +{\kappa^2\over 2\Omega}u_{rn}=-{im\over r}\eta_n,
\label{u2}\label{eq:fluida}\\
&&-i\tomega u_{zn}=-{\eta_n\over h},\label{eq:fluidb}\\
&&-i\tomega {\eta_n\over c_s^2}+
\left({d\over dr}\ln r\Sigma +{n\mu\over r}\right)u_{rn}+
{\mu\over r}u_{r,n-2}+{d\over dr} u_{rn}\nonumber\\
&&\qquad\qquad\qquad +{im\over r}u_{\varphi n}
-{n\over h}u_{zn}=0,
\label{eq:fluid2}\ea
where 
\be
\mu\equiv  {d\ln h/d\ln r}.
\ee
We note here that for the isothermal discs we consider, \mbox{$\mu\neq 0$}. Eliminating $u_{\varphi n}$ and $u_{zn}$ from eqs.~(\ref{u1})-(\ref{eq:fluid2}), we have
\ba
&&{d\eta_n\over dr}={2m\Omega\over r\tomega}\eta_n-{D\over\tomega }iu_{rn}
\nonumber\\
&&\qquad\quad +{\mu\over r}\bigl[n\eta_n
+(n+1)(n+2)\eta_{n+2}\bigr],\label{eq:dwndr}\\
&&{du_{rn}\over dr}=-\left[{d\ln(r\Sigma)\over dr}+
{m\kappa^2\over 2r\Omega\tomega}\right]u_{rn}+{1\over i\tomega}
\left({m^2\over r^2}+{n\over h^2}\right)\eta_n\nonumber\\
&&\qquad\quad +{i\tomega\over c_s^2}\eta_n
-{\mu\over r}(n u_{rn}+u_{r,n-2}),
\label{eq:durndr}
\ea
where we have defined
\be
D\equiv \kappa^2-\tomega^2=\kappa^2-(\omega-m\Omega)^2.
\label{eq:D}
\ee
If we only consider the $n=0$ term in the expansion (\ref{eq:expand}), then $u_z=0$ and eqs. (\ref{eq:dwndr})-(\ref{eq:durndr}) reduce to the dynamical equations for 2D discs. However,  since $\mu\neq 0$, eq.(\ref{u1})-(\ref{eq:fluid2}) are an infinite set of coupled ordinary differential equations and the $n=0$ component is coupled to the $n=2,4,\cdots$ components, giving rise to nontrivial vertical structure.  In our linear calculation below, we will truncate the Hermite series at $n=2$ and only include the $n=0$ and $n=2$ components.

\section{Setup and Method of Solution}\label{methods}

We consider the following Gaussian density bump in the disc:
\be
\Sigma / \Sigma_0=1+\chi \exp\Big[-\frac{(r-r_0)^2}{2\sigma^2}\Big ],
\label{eq_surfd}
\ee
where $r_0$ and $\sigma$ are the position and the width of the bump.
We choose the bump parameters 
\be
\chi=0.25, \quad 
\sigma/r_0=0.05,
\ee
and normalize $r_0=1$. The sound speed is given by $c_s/(r_0\Omega_0)=0.1$, 
where $\Omega_0$ is the Keplerian orbital frequency at $r_0$.
The corresponding surface density and vortensity profiles are 
shown in Figure \ref{fig:dens}.

Equations (\ref{u1})-(\ref{eq:fluid2}) are transformed into an eigenvalue problem and solved using two different methods and numerical
codes:

{\it Method (1)}: We adapt the code (MODINT) originally developed by
\citet{TP99} in their study of the accretion-ejection instability in
2D magnetized discs. We add the equations associated with the $n=2$
components in the code, but the method is otherwise similar.
The radial direction is discretised on a grid
logarithmically spaced with a resolution of $255$ points. The density
is defined at the centre of the grid cells, whereas the velocities are
defined at the cell boundaries. Although the problem is now 3D, it is
simpler than the one studied in \citet{TP99}, as there is no magnetic
field. This allows us to solve the set of equations on the real
integration axis, which give more accurate eigenvectors. 

{\it Method (2)}: We use the relaxation method to solve this two-point boundary eigenvalue problem as detailed in \citet{PTV92}. 
The computation cost is relatively low in this method and we can afford higher resolution. Typically, we use 800 uniform grid points in our calculation. 
 
Since the Rossby waves are concentrated around the density bump, we
choose for both codes the inner boundary at 
$r_{\rm in}/r_0=0.4$ and the outer boundary at $r_{\rm out}/r_0=1.6$. 
The two codes give the same results when the same null 
radial velocity boundary conditions are adopted.

\section{Results of Linear Calculations}\label{results}

\begin{figure}
\begin{center}
\includegraphics[width=\linewidth,trim=1cm 11.5cm 2cm 10.5cm,clip=true]{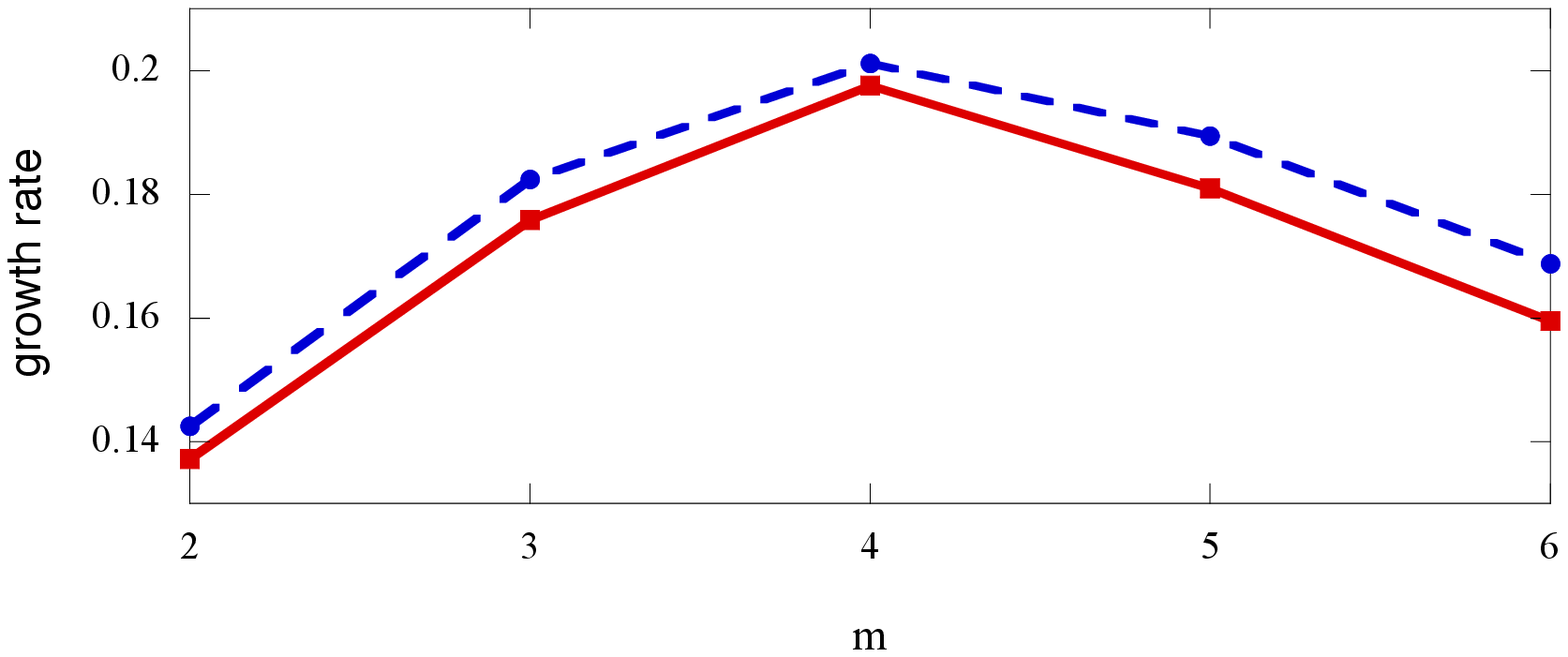}
\end{center}
\caption{The RWI growth rate (in units of $\Omega_0$)
as a function of the azmuthal mode number $m$ 
obtained by linear calculations for 2D ({\it dashed line}) 
and 3D ({\it solid line}) discs.}
\label{fig:lin}
\end{figure}

As noted above, we have implemented two types of boundary conditions
to solve the linear eigenvalue problem. The radiative boundary condition is preferred, since it corresponds to the spontaneous growth of perturbations around the
density bump without energy input from the regions outside the bump. But whatever the boundary conditions, our calculations show that for the choice of disc parameters and computation domain adopted in this paper, the complex eigenvalues $\omega$ are almost identical.
This can be understood since only a small amount of wave energy is leaked
out of the Rossby zone as outgoing (away from the density bump) density waves.

Figure \ref{fig:lin} shows the linear growth rate of the 3D Rossby
mode trapped around the density bump as a function of the azimuthal
wavenumber $m$. The 2D result, obtained by including only the $n=0$
terms in the expansion (\ref{eq:expand}), is also shown for
comparison.  We see that the 3D growth rates are only slightly smaller
than the 2D ones under the same conditions (see Section 4.1 below). 
In all cases, we find that the real part of the mode frequency $\omega_r$ close to $m\Omega_0$, with
$\omega_r/m\Omega_0=0.974$ for $m=2$ and $0.986$ for $m=6$.

\begin{figure*}
\begin{center}
\subfloat[Radial velocity]{\label{fig:eigen-a}\includegraphics[width=0.4\textwidth]{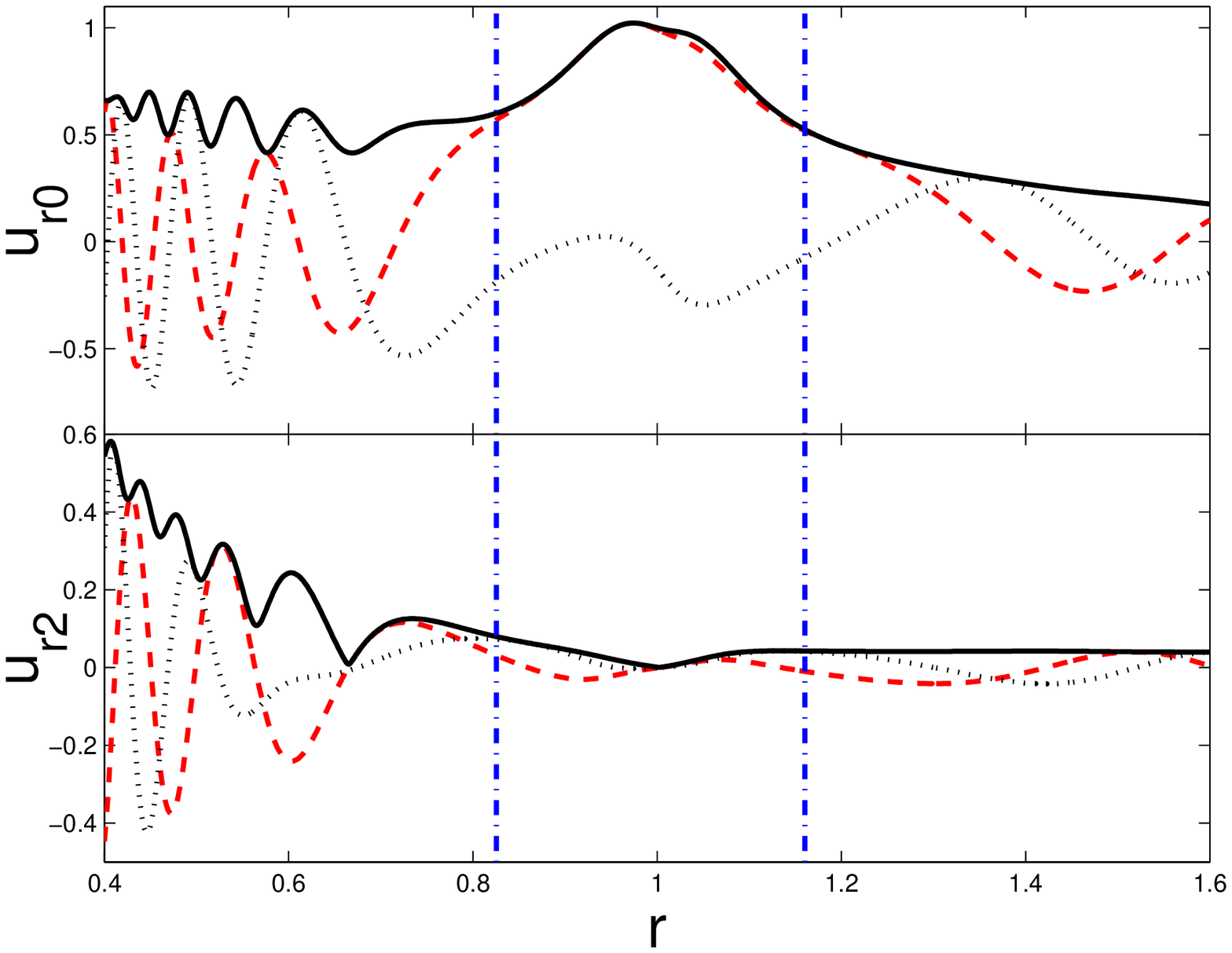}}
\subfloat[Azimuthal velocity]{\label{fig:eigen-b}\includegraphics[width=0.4\textwidth]{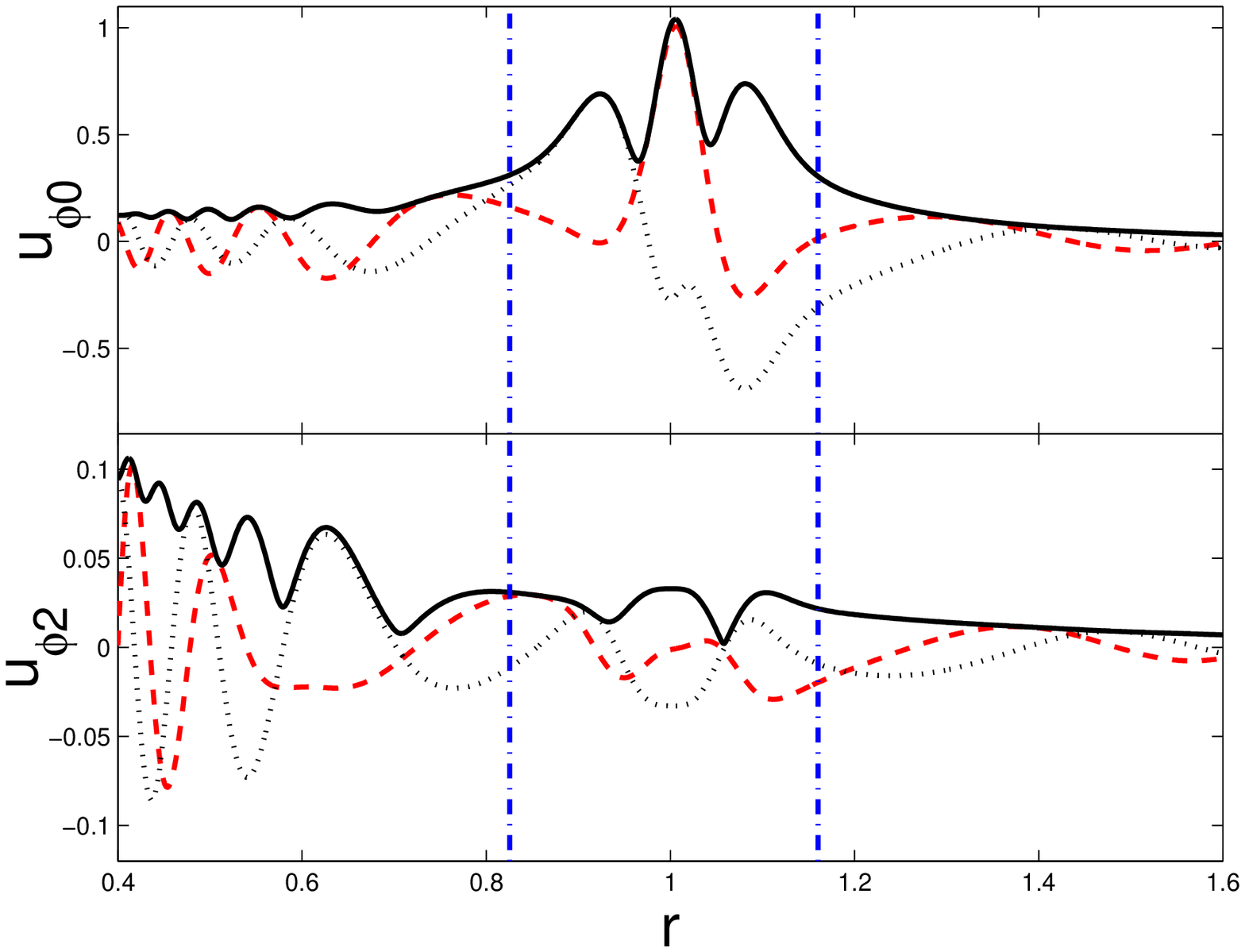}}\\
\subfloat[Densitiy]{\label{fig:eigen-c}\includegraphics[width=0.4\textwidth]{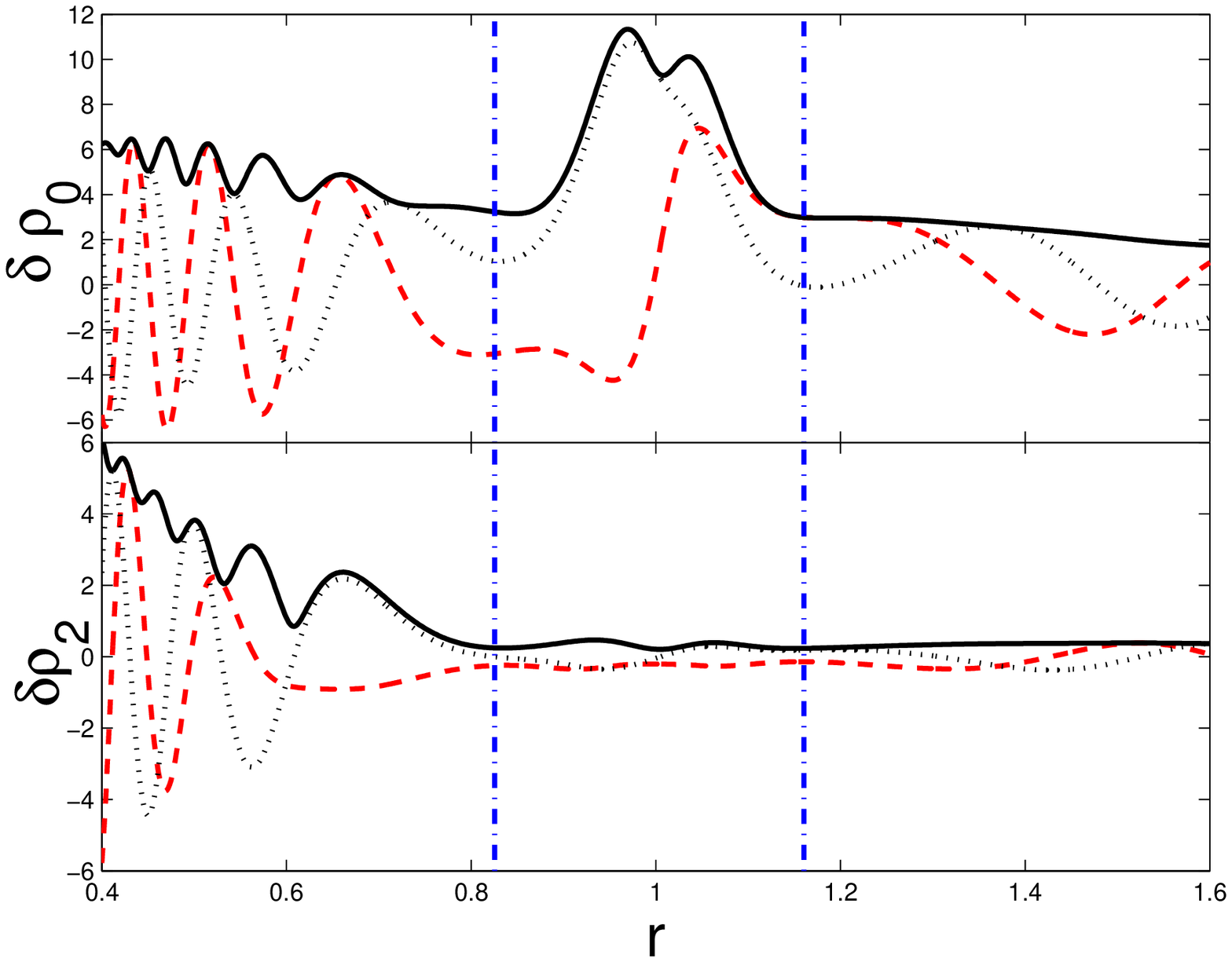}}
\subfloat[Vertical velocity]{\label{fig:eigen-d}\includegraphics[width=0.4\textwidth,trim=0cm -0.8cm 0cm 0cm,clip=true]{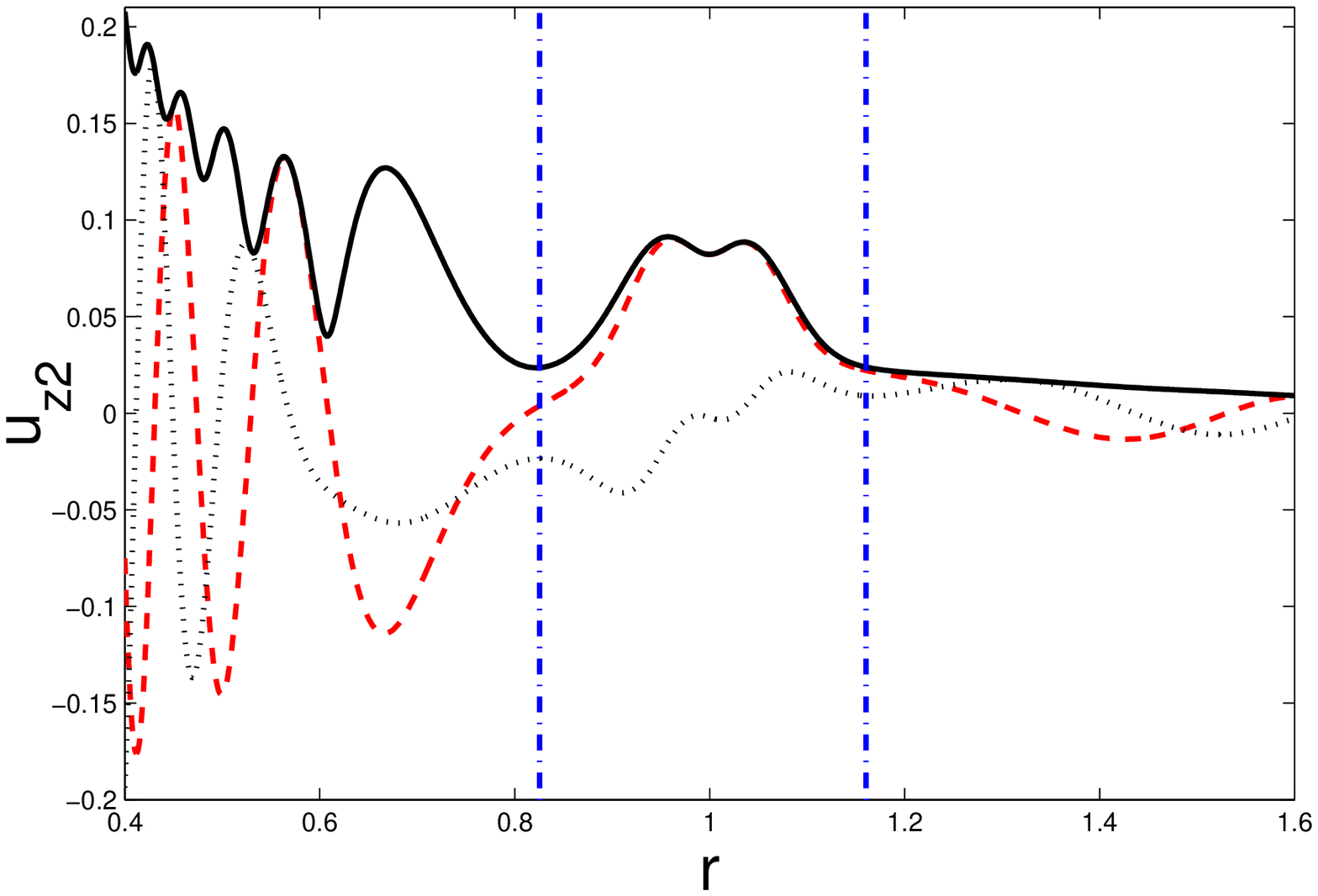}}
\end{center}
\caption{Eigenfunctions of the $m=4$ mode: The dotted line shows the real
  part, the dashed line the imaginary part, and the solid line
  the absolute value. The vertical dot-dashed lines give the
  positions of the inner and outer Lindblad resonances. The radius is in 
units of $r_0$, see also section \ref{results}.}
\label{fig:eigen}
\end{figure*}

Figure \ref{fig:eigen} depicts the eigenfunctions of the $m=4$ mode. For each variable, we show the $n=0$ and $n=2$ components, except for the vertical velocity which has a nul $n=0$ component. The amplitudes of the eigenfunctions are normalised so that the maximum value of the radial velocity perturbation $|u_{r0}|$ equals unity (this maximum occurs at $r\simeq r_0$). Note that the Rossby mode is confined around the corotation radius (where $\omega_r=m\Omega$), which is close to the density bump. But the mode can leak out as spiral density waves inside the inner Lindblad resonance (where $\omega-m\Omega=-\kappa$) and outside the outer Lindblad resonance (where $\omega-m\Omega=\kappa$). [See, e.g., \citealt{TSA08} and \citealt{LAI09} for discussion on the wave zones for Rossby waves and density waves.] The $n=0$ spiral wave (as a function of $r$) satisfies the WKB amplitude relation for $\eta_0$ \citep{TSA08},
\be
|\eta_0(r)|\propto \left({c_s^2|D|\over r^2\Sigma^2}\right)^{1/4}
\ee
and similar relations for $u_{r0}$ and $u_{\varphi 0}$.
The $n=2$ component is driven by the coupling with the $n=0$ component. The vertical velocity has a small but non-zero amplitude, and as this is a pure $n=2$ component, its amplitude increases with height whereas the other variables ($u_r$, $u_\varphi$ and $\delta\rho$) are dominated in the Rossby wave region by the vertically constant component ($n=0$). This means that the vertical velocity component can not be neglected when the flow pattern is of importance, as for instance for the study of the concentration of dust in the vortices. In other words, the thin disc approximation do not capture the full structure of the instability (see also Section 5).

\subsection{Angular momentum flux}\label{discussion}

\begin{figure}
\begin{center}
\includegraphics[width=0.4\textwidth]{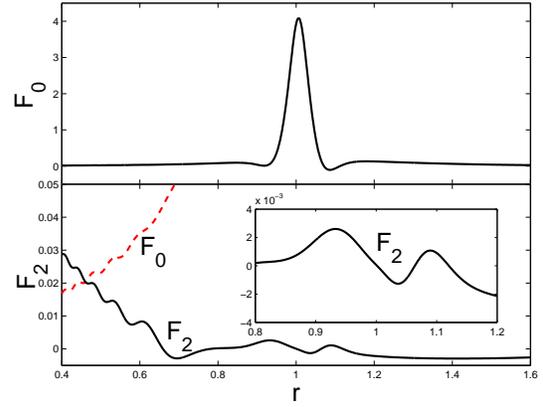}
\end{center}
\caption{Angular momentum flux as defined by eq.~(28) for
  the $n=0$ (upper panel) and $n=2$ (lower panel) components of the
  $m=4$ mode. The radius is in units of $r_0$ and the flux in units of
  $\Sigma_0r_0^4\Omega_0^2$. In the lower panel, the dashed line shows
  $F_0$, and the insert shows a blowup of $F_2$ around the density
  bump.}
\label{fig:flux}
\end{figure}

A useful way to understand the RWI is to examine the angular 
momentum flux carried by the wave. The time averaged transfer rate
of the $z$-component of angular momentum across a cylinder of radius $r$
is given by (see, e.g., \citealt{LK72,GT79,TTW02})
\be
F(r)=\Bigl\langle r^2 \int_{-\infty}^\infty\! dz\int_0^{2\pi}\! d\varphi\,
\rho_e(r,z) u_r(r,\varphi,z,t)u_\varphi(r,\varphi,z,t)\Bigr\rangle,
\ee
where $\langle \rangle$ stands for the time average.
Using $u_r(r,\theta,z,t)=\rmRe\,[{u_r}_{n}H_n(Z) e^{i(m\varphi-\omega t)}]$, 
$u_\varphi(r,\varphi,z,t)=\rmRe\,[{u_\varphi}_{n}H_n(Z)
e^{i(m\varphi-\omega t)}]$, we find that the angular momentum flux
associated with the $n$ component of the $m$ mode is (see \citealt{ZL06})
\be
F_n(r)=n!\,\pi r^2\Sigma\,\rmRe \,(u_{rn}u_{\varphi n}^*).
\label{eq_flux}
\ee

Figure \ref{fig:flux} shows the angular momentum fluxes for the $n=0$ and $2$ components of the $m=4$ mode. Around the density bump, the flux is dominated by the $n=0$ component. Thus it is not surprising that the 3D linear growth rate is close to the 2D value (see Figure~2). Indeed, the growth of the RWI arises from the positive $F_0$ around the corotation (close to the density bump): since the perturbation inside (outside) the corotation carries negative (positive) angular momentum, a net outward angular momentum transfer across the corotation induces mode growth.

The finite $n=2$ wave component arises from its coupling to the $n=0$ component. As detailed in \citet{ZL06} (see their section 6.2), $n\ge 1$ waves propagating across the 
corotation are strongly attenuated, and the corotation acts as a sink for waves with $n\ge 1$ (see also \citealt{PAP85,K03,LGN03,LAB09}).
This is consistent with the behavior of $F_2$ around the corotation: we see from the insert of Figure~\ref{fig:flux} that $F_2$ changes sign from positive to negative across the corotation radius.
This explains our numerical result (see Figure~2) that the 3D RWI growth rate is smaller than the 2D growth rate.

\section{Comparison to Numerical Simulations}\label{simulations}

We have performed full numerical simulations with the same disc configuration and parameters as those used in our linear calculations. The comparison to simulations is important 
for different reasons:
\begin{enumerate}
\item In our linear calculations, we included only the first two
  elements ($n=0,2$) of the Hermite polynomial decomposition, the higher 
  order terms being neglected. The non-linear simulations can confirm if this
  approximation is correct.
\item The previous simulations of (Meheut et al., 2010) showed
  unexpected vertical structure in the Rossby waves. New simulations
  with the same setup as the linear analysis can confirm if the
  vertical structure is correctly handled and if the simulations are
  not altered by the numerical methods or boundary conditions.
\end{enumerate}

\subsection{Numerical methods}

The numerical methods used for this study are very similar to the one
of Meheut et al., 2012. We used MPI-AMRVAC \citep{KEP11} with the total variation diminishing Lax-Friedrich scheme with a third order accurate Koren limiter. The grid is cylindrical with
$r/r_0$ in the range of $[0.4, 1.6]$, $z/r_0$ in the range of $[0, 0.6]$,
and full azimuthal range $[0,2\pi]$ is covered.
This vertical extension of the grid corresponds to $6$ scale height at $r_0$. A high vertical extension of the grid is needed to resolve the vertical structure of the disc at
the outer edge of the grid. The grid is fixed during the simulation but the initial resolution is determined to fit the regions of high density gradient. The base resolution is $(128,32,32)$ and up to four level of refinement are allowed, reaching a resolution of $(2048,512,512)$ near $r_0$.  The boundary conditions are the same as Meheut et al. (2012) with a null radial velocity at the boundaries. This boundary condition is slightly different from the radiative boundary condition. However, as discussed in Section 4, the two different boundary conditions produce nearly identical linear mode frequency and growth rate.

In our simulations, the initial midplane density profile is given by
\be
\rho_M(r)=\frac{1}{\sqrt{2\pi}h}\left\{ 1+\chi\exp\left[{-\frac{(r-r_0)^2}
{2\sigma^2}}\right]\right\}.
\ee
The vertical profile of the density is then chosen to achieve hydrostatic 
equilibrium:
\be
\rho_e(r,z)=\rho_M(r)\exp\left[{\frac{GM}{c_s^2}\Big(\frac{1}{\sqrt{r^2+z^2}}-\frac{1}{r}}\Big)\right].
\ee
This gives a surface density profile very similar to eq.~24. The azimuthal velocity is determined by force balance in the radial direction. 

The small difference from the linear analysis resides in the upper region of the simulation where the density reaches very low values. In the simulations, there is a floor value for the density. To check if this difference is of importance we have performed some tests, one with a floor density of one tenth of its initial value and one where the height of the numerical box has been doubled has been doubted. We find a negligible difference in the resulting growth rates.

One simulation has been performed without any perturbation to check that the initial state is indeed an equilibrium state. For the other simulations we add small perturbations in the radial velocity:
\be
\frac{v_r}{r_0\Omega_0}=\epsilon \sin(m\varphi)\exp\left[-\frac{(r-r_0)^2}{2\sigma^2}\right],
\label{eq_vr}
\ee
with $\epsilon\sim5\times 10^{-4}$ and $m=2,3,\cdots,6$.

\subsection{Results of the simulations}

\begin{figure}
\begin{center}
\includegraphics[height=\linewidth,angle=90,trim=1cm 3.5cm 2cm 1.5cm,clip=true]{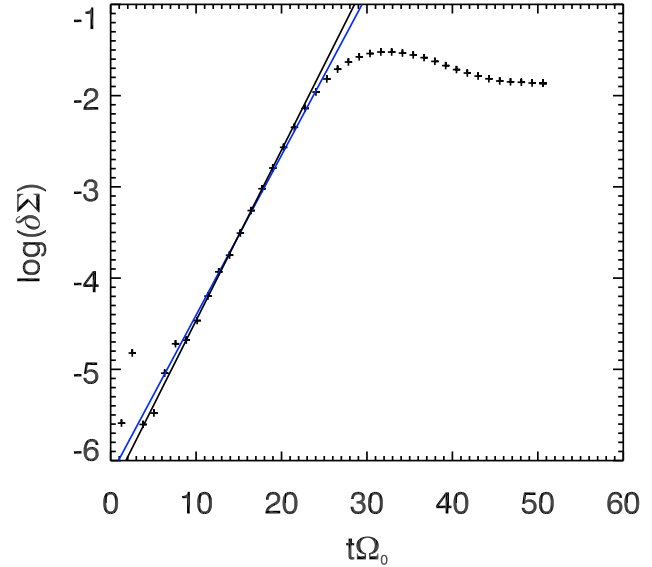}
\end{center}
\caption{The amplitude of the density perturbation (on a logarithmic scale) as a function of time (in units of $1/\Omega_0$) in the $m=3$ simulation. The $y$-axis shows the logarithm of the surface density perturbation $\delta\Sigma$, where $\delta\Sigma={\rm max}_{r,\varphi}|\Sigma-\langle\Sigma\rangle_\varphi|$.
The two solid lines represent linear fits to the exponential
growth phase of the RWI. The difference in the fitted linear growth rates 
is about 5\%.}
\label{fig:simlingrowth}
\end{figure}

\begin{figure}
\begin{center}
\includegraphics[height=8cm,trim=2.5cm 0.5cm 0cm 1.2cm ,clip=true]{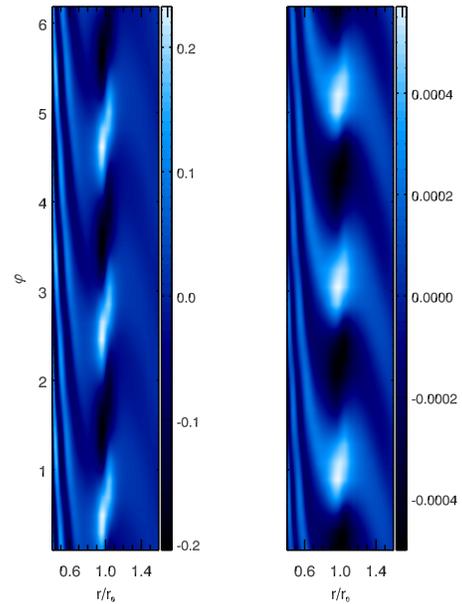}
\end{center}
\caption{Perturbed density (\emph{left}) and radial velocity (\emph{right}) in the midplane at $t\Omega_0\sim18$ in the $m=3$ simulation.}
\label{fig:vrmid}
\end{figure}

\begin{figure}
\begin{center}
\includegraphics[scale=0.5,trim=1.5cm 11.5cm 1cm 10.5cm,clip=true]{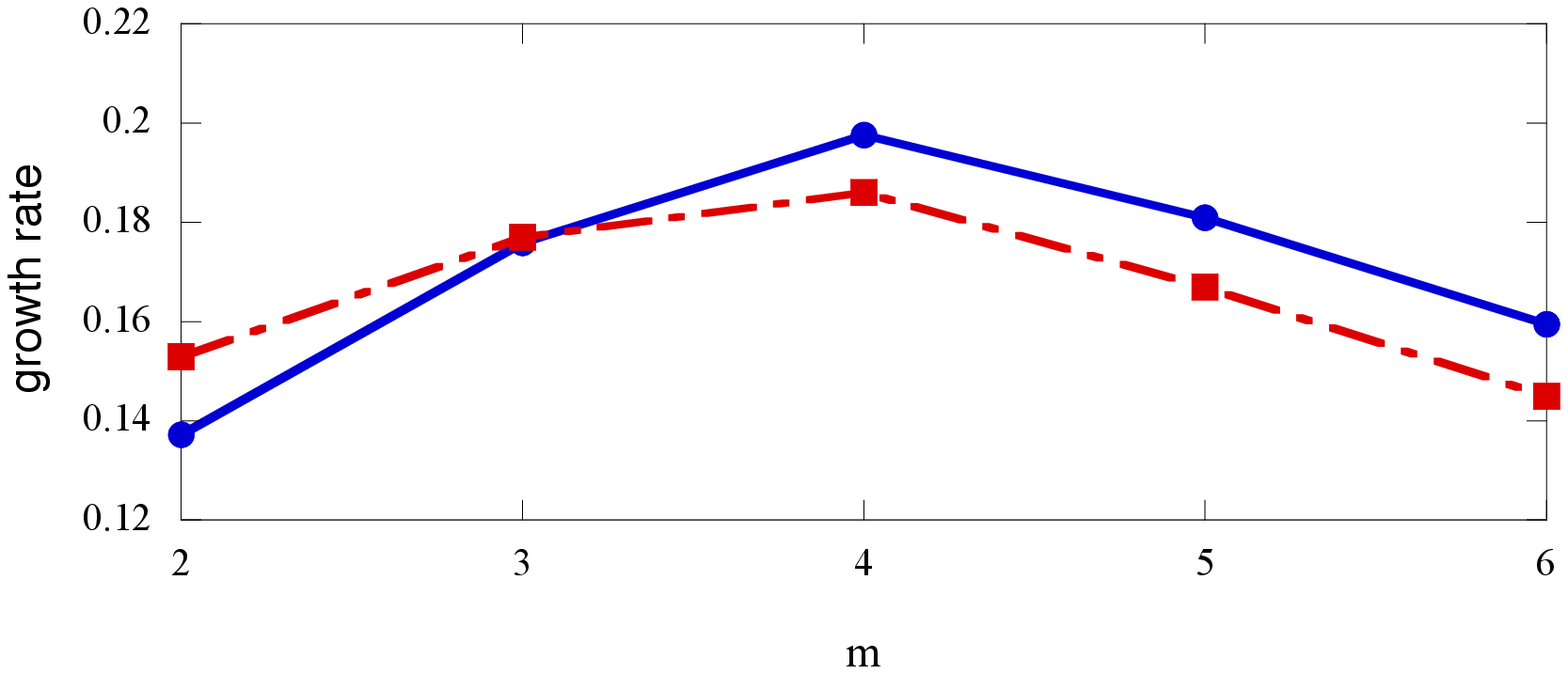}
\end{center}
\caption{Growth rate obtained for different azimuthal modes with the
  $2$ methods: the linear analysis (\emph{solid line}) and the non-linear simulations (\emph{dashed line}).}
\label{fig:sim}
\end{figure}

Starting from the initial perturbation (31) for a given $m$, the simulation follows the exponential growth of the instability and the saturation phase on timescale of a few tens of rotations. The exact timescale to reach saturation depends on the azimuthal mode number $m$.  A fit of
the exponential growth can give the growth rate $\gamma$ of the RWI. An example (for $m=3$) is shown in Figure \ref{fig:simlingrowth} and the density and radial velocity perturbation during the linear phase ($t\Omega_0\sim18$) are shown in Figure \ref{fig:vrmid}. The azimuthal mode number ($m=3$) is clearly seen in this figure.

Figure \ref{fig:sim} shows the linear RWI growth rates obtained from our 3D simulations for different $m$'s. We point out here that the estimated growth rate depends on the  defined position of the beginning and the end of the linear phase. The growth rates of the different $m$'s are then given with an uncertainty of $5$ to $10\%$ due to the difficulty to find the best fit for the linear growth. The results are compared with the mode growth rates from our linear perturbation calculations. We see that the numerical growth rates are in good agrement with the linear perturbation theory results. For the $m\ge 4$, the non-linear simulations give slightly smaller growth rates as compared to the linear perturbation results. This could be due to the higher numerical viscosity on the smaller-scale modes or to the higher ($n>2$) terms in the Hermite decomposition that have not been considered in the linear approach that would be more important for higher $m$.

\begin{figure}
\begin{center}
\includegraphics[width=\linewidth,trim=0.5cm 6cm 1.5cm 7cm ,clip=true,angle=90]{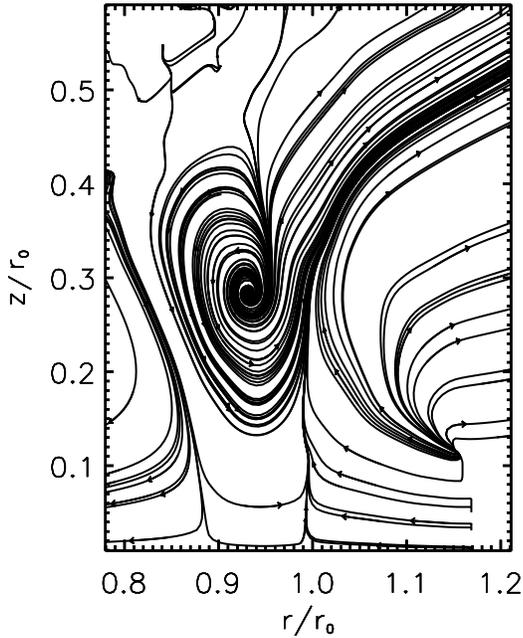}
\end{center}
\caption{Velocity streamlines in a vertical frame at $t\Omega_0\sim 18$.}
\label{fig:stream}
\end{figure}

Some velocity streamlines in a vertical frame of the disc are plotted
in Figure \ref{fig:stream} showing one of the vertical roll structures we are studying in this paper.

\begin{figure*}
\begin{center}
\subfloat[Radial velocity]{\label{fig:eigensim-a}\includegraphics[width=0.3\textwidth,trim=1cm 1.4cm 2cm 1.5cm,clip=true,angle=90]{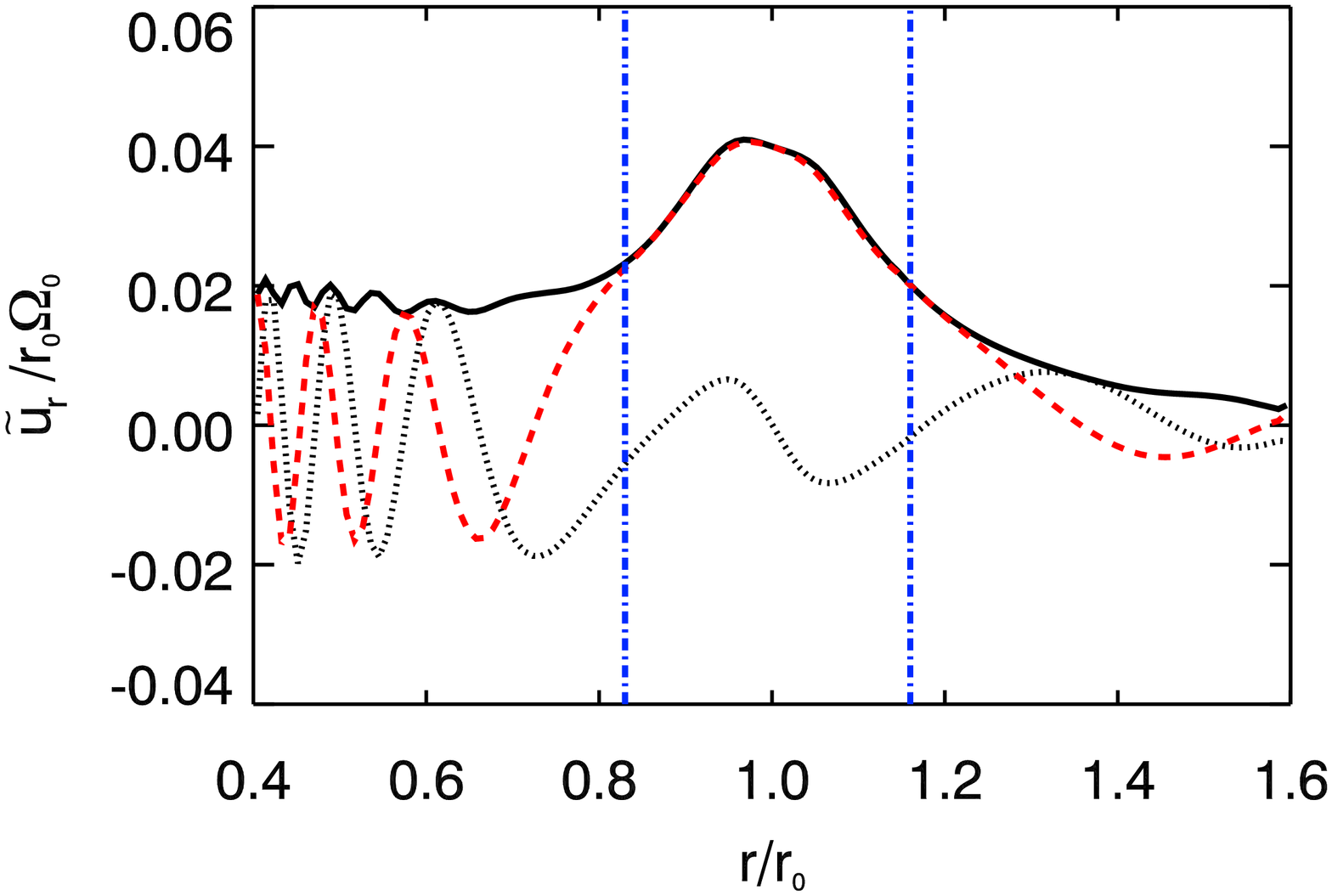}}
\subfloat[Azimuthal velocity]{\label{fig:eigensim-b}\includegraphics[width=0.3\textwidth,trim=1cm 1.4cm 2cm 1.5cm,clip=true,angle=90]{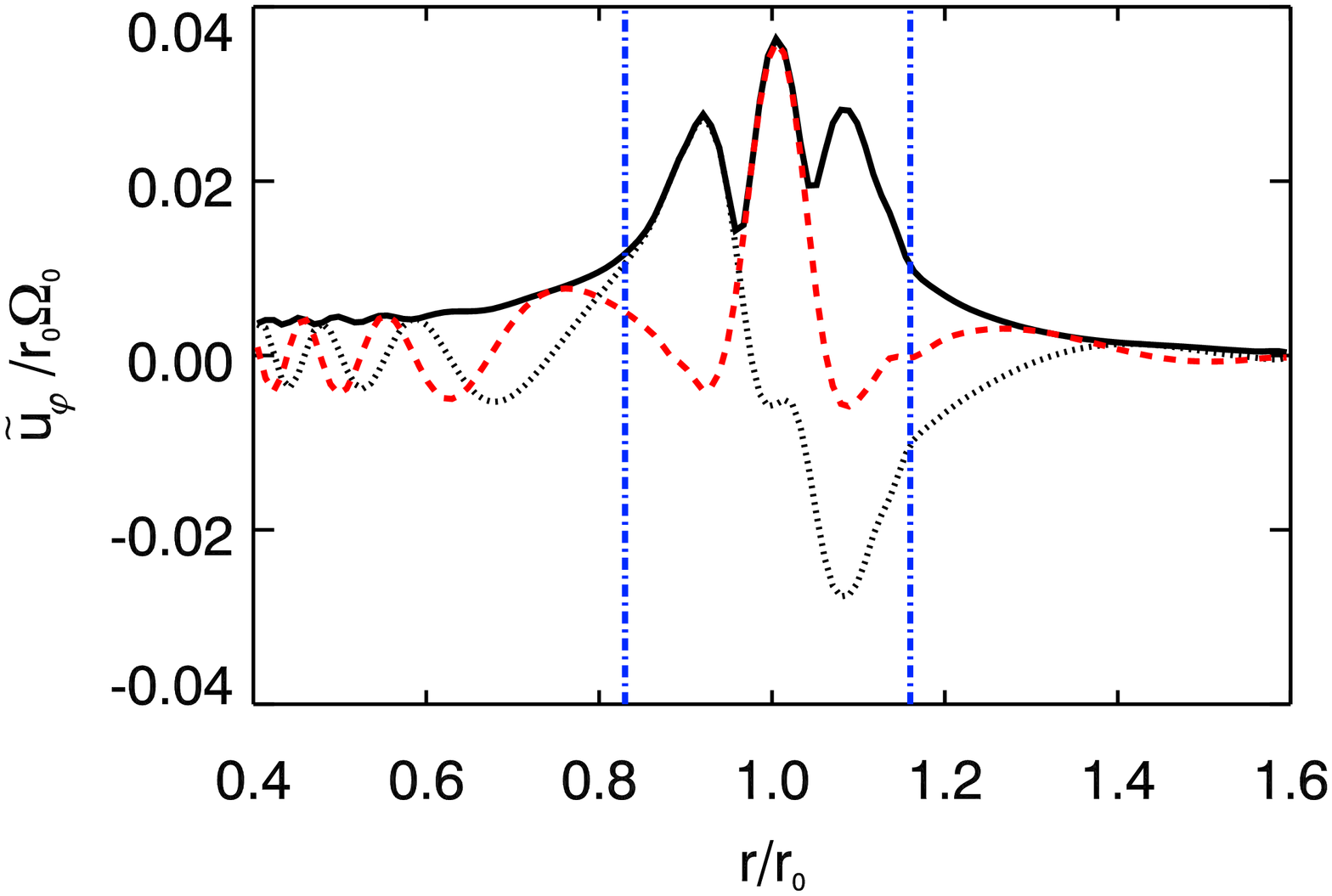}}\\
\subfloat[Densitiy]{\label{fig:eigensim-c}\includegraphics[width=0.3\textwidth,trim=1cm 1.4cm 2cm 1.5cm,clip=true,angle=90]{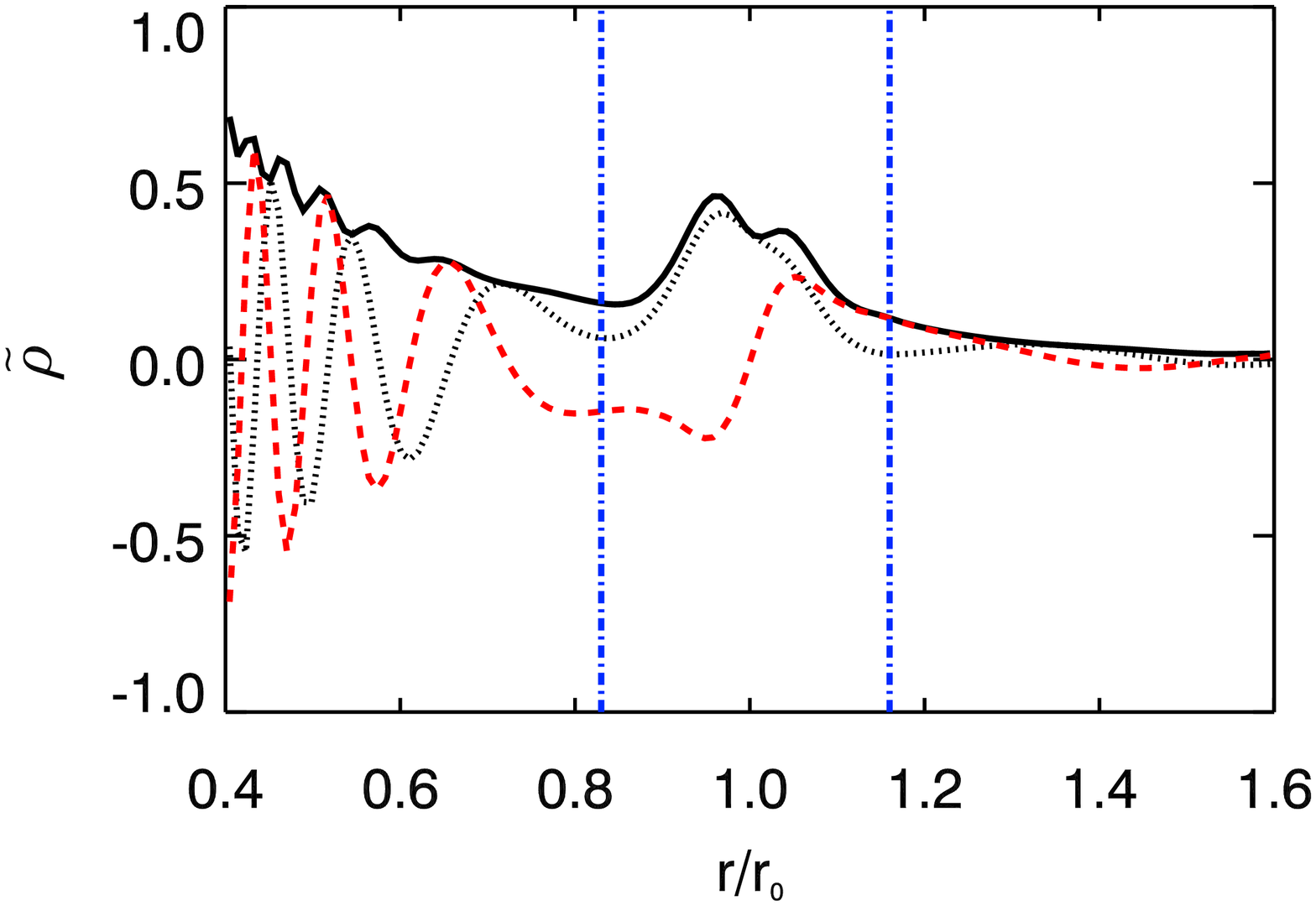}}
\subfloat[Vertical velocity]{\label{fig:eigensim-d}\centering\includegraphics[width=0.3\textwidth,trim=1cm 1.4cm 2cm 1.5cm,clip=true,angle=90]{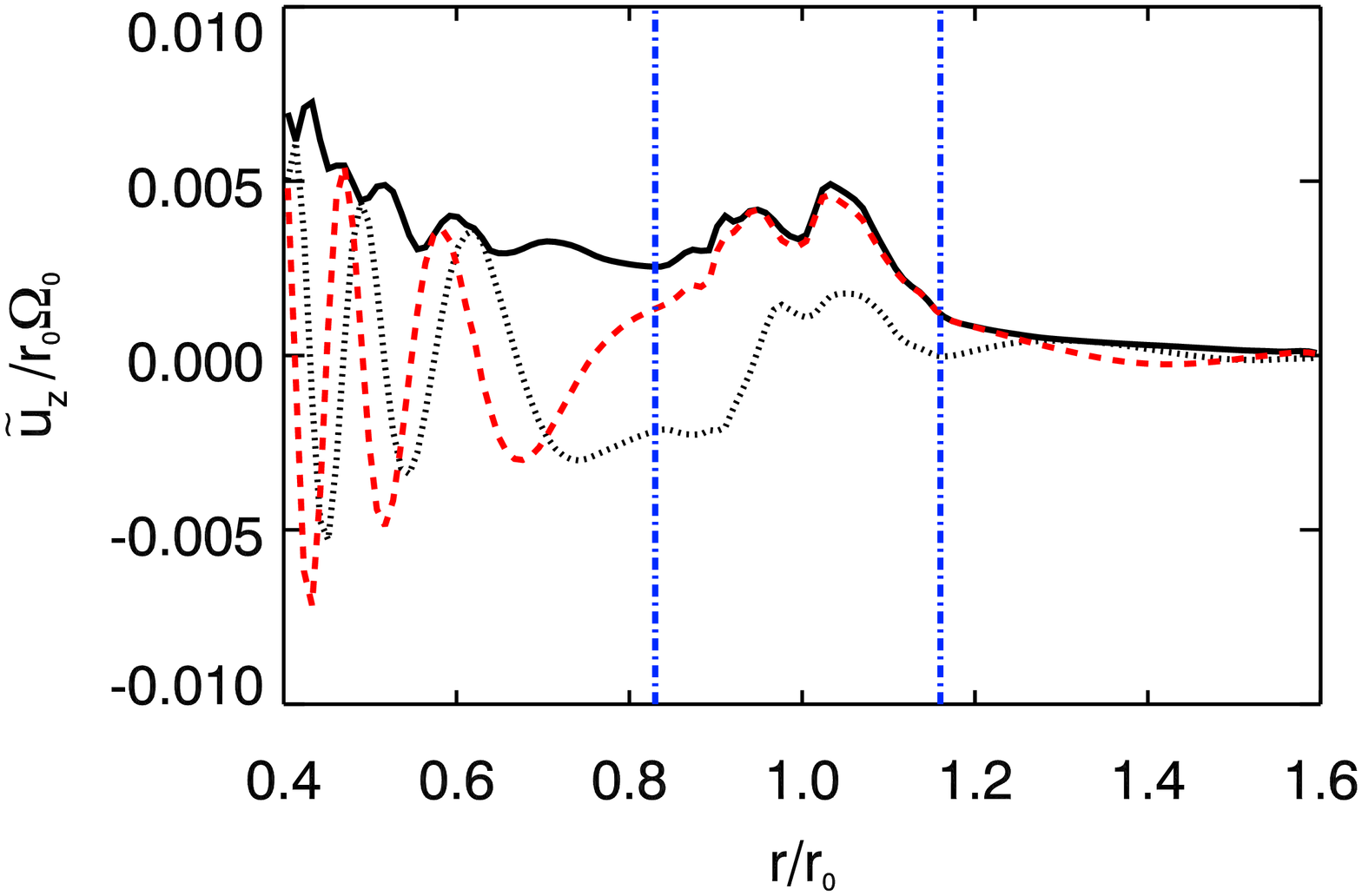}}
\end{center}
\caption{Numerical 'eigenfunctions' of the $m=4$ mode obtained with the non-linear code at
$t\Omega_0\sim 18$: 
The dotted line is the real part, the red
  dashed line is the imaginary part, and the solid line is the
  amplitude of the $m=4$ element of the azimuthal Fourier transform of the physical quantities.}
\label{fig:eigensim}
\end{figure*}

\begin{figure}
\begin{center}
\includegraphics[height=\linewidth,trim=1cm 1.4cm 2cm 2.5cm,clip=true,angle=90]{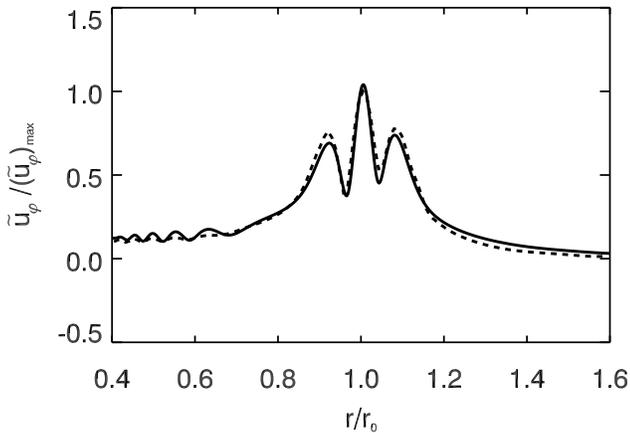}
\end{center}
\caption{Direct comparison of the linear (solid line) and numerical (dashed line) approaches for the azimuthal velocity component. The same normalisation has been used for the two approaches.}
\label{fig:comparison}
\end{figure}

To compare the flow velocity and density structures obtained from our 3D simulations with the linear eigenfunctions, we can apply Fourier transform in the azimuthal direction to our numerical flow outputs. Figure \ref{fig:eigensim} gives an example (for the $m=4$ mode at time
$\Omega_0 t\simeq 18$) of these numerically determined ``eigenfunctions''. 
Note that the density, radial and azimuthal velocities are evaluated at the disc midplane, whereas the vertical velocity is plotted slightly above the midplane. 
This figure should be compare with Figure~\ref{fig:eigen}, where the same eigenfunctions from the linear perturbation theory are shown. Here we have not separated the different vertical
components but one can see that the $n=0$ component dominates except obviously for $u_z$. On Figure \ref{fig:comparison}, we have plotted the amplitude of the azimuthal velocity eigenfunction obtained with the two methods and with the same normalisation, so they can be directly compared. However for the linear calculation only $n=0$ is taken into account whereas the multiple components of the vertical structure have not been separated in the numerical simulations. The good agreement between 
our numerical simulations and linear analysis indicates that the simulations can correctly describe the RWI. The complicated radial structure of the mode in the corotation region coupled with the vertical structure is responsible for the complexity of the flow, as plotted in Figure \ref{fig:stream}.

\section{Conclusion}\label{conclusion}

In this paper we have carried out linear analysis of the Rossby Wave Instability (RWI) in 3D stratified isothermal discs. Our linear calculations show that the vertical velocity inside the Rossby vortices, that was obtained in previous numerical simulations \citep{MEH10} is expected when the disc scale height varies with the disc radius. This vertical velocity is of crucial importance for understanding the concentration of dust inside the vortices and the formation of planetesimals. Detailed comparison with 3D numerical simulations show that the simulations of
Meheut et al.~(2010, 2012) can correctly handle the RWI in 3D. Hence these simulations are robust tools to study the non-linear and long-term evolution of this instability. We have also shown that
the linear growth of Rossby waves is only slightly affected by the disc vertical stratification, with similar mode growth rates in 3D as in 2D. The small reduction of the 3D mode growth rate is due to 
the absorption of the wave component with vertical structure at the corotation resonance.

We have also calculated the angular momentum flux carried by the Rossby vortices. This shows that the growth of the instability is due to the exchange of angular momentum between two Rossby waves on each
side of the density bump. The angular momentm transfer tends to reduce the initial density bump. This is an important feature of the RWI as it may lead to the transfer of angular momentum through the
dead zone of protoplanetary discs. To fully understand this process and the competition with the bump formation process, one will need a self-consistent simulation including a dead-zone inside a ionised disc. In the mean time, analytical model of these processes as the one presented here, can give a first insight of this dynamical evolution.

\section*{Acknowledgments}
This work has been supported in part by AST-1008245 and the Swiss National Science Foundation. C. Y. thanks the support from National Natural Science Foundation of China (Grants 10703012, 11173057 and 11033008) and Western Light Young Scholar Program of CAS.

\bibliographystyle{mn2e}


\label{lastpage}
\end{document}